\documentclass[twocolumn,showpacs,preprintnumbers,amsmath,amssymb,superscriptaddress]{revtex4-1}
\usepackage{amsmath}
\usepackage{graphicx}
\usepackage[T1]{fontenc}
\usepackage{graphicx}
\usepackage{dcolumn}
\usepackage{bm}
\usepackage{xspace}
\usepackage{epstopdf} 

\newcommand{\bra}[1]
{
\langle  #1 |
}
\newcommand{\ket}[1]
{
|#1 \rangle
}

\newcommand{\expect}[1]
{
\langle #1  \rangle
}

\begin{document}
\title{Mapping spin-polarised transitions with atomic resolution
}
\author{P. Schattschneider}
\affiliation{Institut f\"ur Festk\"orperphysik, Technische
Universit\"at Wien, A-1040 WIEN, Austria}
\affiliation{University Service Centre for Electron Microscopy,  Technische Universit\"at Wien, A-1040 WIEN, Austria}
\author{B. Schaffer}
\affiliation{SuperSTEM, STFC Daresbury Laboratories, Keckwick Lane, Warrington, WA4 4AD, UK}
\affiliation{Kelvin Nanocharacterisation Centre, SUPA School of Physics and Astronomy, University of Glasgow, Glasgow, G12 8QQ, Scotland, UK}
\author{I. Ennen} 
\affiliation{University Service Centre for Electron Microscopy,  Technische Universit\"at Wien, A-1040 WIEN, Austria}
\author{J. Verbeeck}
\affiliation{EMAT, University of Antwerp, Groenenborgerlaan 171, 2020 Antwerp, Belgium}



\begin{abstract}
The coupling of  Angstrom-sized electron probes with  spin polarised  electronic transitions shows that the inelastically scattered  probe electron  is in a mixed state containing electron vortices with non-zero orbital angular momentum.  These electrons create an  asymmetric intensity distribution in energy filtered diffraction patterns, giving access to  maps of the  magnetic moments with atomic resolution. A feasibility experiment shows  evidence of the predicted effect. Potential applications are column-by-column maps of magnetic ordering, and the creation of Angstrom-sized free electrons with orbital angular momentum  by inelastic scattering in a thin ferromagnetic foil.

\end{abstract}

\maketitle
\section{Introduction}
With the availablity of electron vortices of sub-nm scale in the transmission electron microscope (TEM)~\cite{VerbeeckNature2010,VerbeeckAPL2011,SchattUM2012a,VerbeeckUM2012a} and its theoretical description\cite{Bliokh2007,SchattUM2011,BliokhPRL2011} many  potential applications come within reach, ranging from the transfer of angular momentum to nanoparticles, over utilisation of the intrinsic magnetic moment of vortex electrons to the probing of chirality\cite{MullerNature}.  Indeed, chiral electronic transitions   were the first application of electron vortices in energy loss magnetic chiral dichroism (EMCD)~\cite{VerbeeckNature2010}. 

The discovery of  EMCD~\cite{SchattschneiderNature2006} was an unexpected alternative to XMCD (X-ray magnetic circular dichroism) with the convenient side effect that additional information on the investigated material can be obtained simultaneously via standard analytical techniques\cite{RubinoJMR2008,SchattschneiderJAP2008}.  The spatial resolution of this electron microscopic technique    
is now in the nanometre range~\cite{SchattschneiderPRB2008}. A modification of the technique has been shown to be site selective~\cite{MMMsubmitted}. The excellent spatial resolution and the site selectivity are 
important for the study of novel materials such as Heusler alloys\cite{yu06,shamberger09}, nanoparticles or interfaces~\cite{KlieAPL2010}. 
Recent advances in electron microscopy have led to the imaging of condensed matter with subatomic resolution~\cite{NellistScience2004,KrivanekNature2010,MullerNature2009}. 
On this basis it has been speculated that the mapping of spin polarised electronic  transitions - and thus the mapping of spin and orbital polarisation  - on the atomic scale could be feasible in a TEM, applying the EMCD technique\cite{EMCDbook}. An incident plane wave affecting a spin polarised L$_{23}$ transition would break the mirror symmetry of non-magnetic transitions in the scattered wave. This symmetry breaking could  be analysed either with asymmetric  objective apertures or   with a cylinder lens\cite{SchattschneiderUM2009,SchattschneiderPRB2010}. 
However, for technical reasons both approaches are unrealistic in the TEM.

On second thoughts it becomes evident that the inelastic interaction  of an incident electron with a spin polarised electronic transition creates a scattered  electron with topological charge. One can thus apply the  theory of vortex electrons\cite{SchattUM2011} to the outgoing wave field. In a sense, this is a bottom-up application of the original idea of using incident vortex electrons for EMCD. The reason that this works is the generalized reciprocity theorem\cite{FindlayUM2007} that confirms  the equivalence of the incident and the outgoing (also called reciprocal\cite{Kainuma}) electron  for inelastic scattering. 
This observation raises two questions: How can an EMCD signal be detected with Angstrom-sized Scanning TEM (STEM) probes; and can one produce  electron vortices without holographic masks? 

Here, we present a theoretical and numerical analysis of the coupling between an Angstrom-sized STEM probe and an  atom-sized vortex field via a chiral  electronic transition. It is shown that the inelastically scattered  probe  is in a mixed state containing electron vortices with non-zero angular momentum.  These electrons create an  asymmetric intensity distribution in energy filtered diffraction patterns, giving access to  maps of the magnetic moments
 on an atomic  column-by-column basis. A feasibility experiment shows  evidence of the predicted effect. Finally  potential applications are discussed: maps of magnetic ordering with atomic resolution; and the creation of free electrons with orbital angular momentum and a diameter of about 0.1~nm by inelastic scattering in a thin ferromagnetic foil.
 

\section{Theory}
We focus on the model of a thin  (ideally one atom) layer of Fe. In this case the dynamical equation for the propagation of the probe's density matrix is considerably simplified. We give here only the basic equation for the propagator and refer the reader to the relevant literature~\cite{EMCDbook,SchattschneiderPRB2010,DwyerPRB2008,SchattschneiderPRB1999,Allen95}. 

 The inelastic intensity at energy loss $E$ in the exit plane of the specimen  is the diagonal term $\rho({\bf r,r})$ of the  density matrix: 
\begin{eqnarray}
\rho_E({\bf r},{\bf r'})&=&
\int  \int \bar G_{d-z}({\bf r},{\bf x}) \bar G_{d-z}^*({\bf r}',{\bf x}')T_{E}({\bf x},{\bf x}',z,z') \nonumber \\
&& \phi_z^\ast({\bf x})  \phi_{z'}({\bf x}')    d^2x \, d^2x' e^{iq_e(z-z')} dz dz'  .
\label{rhoout2}
\end{eqnarray}
where  $z, z'$ are variables along the optic axis  and $\bf r, x$ are in  planes perpendicular to the optic axis. $G$ is the elastic propagator of electrons in the crystal. $\phi_z$ is the wave function of the incident  electron at depth $z$ 
 and $q_E=k_o -k_i$ is the minimum wave number  transfer in the inelastic interaction, equal to the difference between wave numbers of the outgoing and the  incident electron. 

The inelastic scattering kernel $T_{E}({\bf x},{\bf x}',z,z')$ is characteristic for the electronic transitions creating an energy loss $E$. For thin specimens and  atomic columns without defects  the $z$ integration can be performed in closed form \cite{VerbeeckUM2005}.  For spin polarised dipole transitions the atomic scattering kernel reads~\cite{SchattschneiderPRB2010}
\begin{equation}
T_E({\bf x},{\bf x})=
 \sum_{\mu=-1}^1  |\psi_\mu({\bf x})|^2 \sum_{s= \uparrow, \downarrow}
C_{j \mu}^{\uparrow, \downarrow} n^{\uparrow, \downarrow}.
\label{Imu}
\end{equation}
with
\begin{subequations}
\label{psi2}
\begin{equation}
\psi_{\pm 1}({\bf x})= e^{\pm i\alpha}
\frac{i}{2 \pi} \int_0^\infty \frac{q^2 J_1(q x) \langle
j_1(Q) \rangle_{ELSj}}{Q^3} dq 
\label{psi2a}
\end{equation}
\begin{equation}
\psi_0({\bf x}) = \frac{q_E}{2 \pi} \int_0^\infty \frac{q J_0(q x) \langle
j_1(Q) \rangle_{ELSj}}{Q^3} dq,
\label{psi2b}
\end{equation}
\end{subequations}
with the matrix element  of the spherical Bessel function
$$
\langle
j_1(Q) \rangle=\bra{I}
j_1(Q) \ket{F}
$$
between initial and final target states. $Q^2=q^2+q_E^2$ with the characteristic wave number transfer $q_E=k_0 E/(2 E_0)$.
The  coefficients $C_{j \mu}$ are weighting factors for spin-orbit coupling\cite{TholePRL92,RuszPRB07t,SchattschneiderPRB2010}, and $n^{\uparrow, \downarrow}$ is the spin polarisation of the final state. The spin-orbit coupling of the initial state renders the coefficients $C$ dependent on the total magnetic quantum number $j=l+s$. For the $L_{23}$ edges to be considered  $j=1/2$ or $j=3/2$. The essential property that we will focus on is described by Eq.\ref{psi2a}: It represents the outgoing inelastically scattered wave as an electron vortex with topology $m=\pm 1$ in the form of a Hankel transform that is easily computable from atomic wave functions. The azimuthal phase factor $e^{\pm i \alpha}$ shows that the outgoing probe electron has  orbital angular momentum. It should be mentioned that the probe beam is not spin polarised. It has acquired orbital angular momentum by spin-orbit coupling of the target electrons. 

The propagation of focussed probes through a thin specimen has regained interest in the context of real space STEM~\cite{EtheridgePRL2011}. Even for elastic scattering the problem of propagating a focussed probe - as we shall adopt in the following - through a thin specimen  to the detector poses considerable numerical problems. The inelastic interaction that can take place throughout the specimen  adds another complexity. Therefore Eq.\ref{rhoout2} cannot be solved without approximations, at least with present numerical capacity.  We  restrict the discussion to a model system that allows to analyse the salient features of the inelastic coupling process with an accuracy comparable to the available experimental data.
As such we choose a line of equally spaced atoms with given spin polarisation; we shall calculate the contributions from each transition 
in dipole approximation, discuss the signal from a single atom and finally build  a line profile of the energy filtered signal from the array of atoms.
For this model the elastic  propagators  $G$ in Eq.\ref{rhoout2} collapse into  delta functions
 and we have for the diagonal element (the density)
\begin{equation}
\rho_E({\bf r},{\bf r})=
T_{}({\bf r},{\bf r}) 
 \phi_0^\ast({\bf r})  \phi_{0}({\bf r}).
\label{rhoout3}
\end{equation}
Eq. \ref{Imu} allows us to disentangle the problem: Each transition channel $\mu \in [-1,1]$ can be factorised into a product of wave functions, and we get for the outgoing intensity in channel $\mu$
\begin{equation}
\rho_{j \mu}({\bf r},{\bf r})=
{\bar{ C}_{j \mu}}
\varphi_\mu({\bf r})\varphi_\mu^\ast({\bf r})
\label{4}
\end{equation}
with 
\begin{equation}
{\bar{ C}_{j \mu}}=\sum_{s= \uparrow, \downarrow}
C_{j \mu}^{\uparrow, \downarrow} n^{\uparrow, \downarrow}
\label{Cj}
\end{equation}
and 
\begin{equation}
\varphi_\mu({\bf r})=\psi_\mu({\bf r}) 
 \phi_i^\ast({\bf r}). 
\label{phimu}
\end{equation}
The intensity in the diffraction plane is calculated from the 2D-Fourier transform of Eq.~\ref{4}:
\begin{equation}
\rho_{j \mu}({\bf q},{\bf q})={\bar{ C}_{j \mu}}
 |FT_{\bf r}[\varphi_\mu({\bf r})]|^2
\label{4bb}
\end{equation}
as the trace over the 3 transition channels $\mu$:
\begin{equation}
\rho_{j,E}({\bf q},{\bf q})=Tr_\mu [\rho_\mu({\bf q},{\bf q})]=\sum_\mu \rho_\mu({\bf q},{\bf q}) \, .
\label{4b}
\end{equation}
The trace operator shows formally that the scattered electron  is in a mixed state. With the coefficients $C$ for $L_{23}$ transitions we can compute the outgoing intensity and the corresponding diffraction patterns. Without loss of generality we assume complete spin polarisation for the final target states (as is justified for the L edges of the $3d$ ferromagnets that we will use as a demonstration  example). 

For complete spin  polarisation the values for $\bar C_{j \mu}$ are given in Table~\ref{tab:xxx}~\cite{SchattschneiderPRB2010}.
The image intensity Eq.\ref{rhoout3} will then contain different contributions from the scattering channels for spin up and spin down polarisations. This difference is the basis of EMCD. 

\begin{table}[h]
	\centering
	\begin{tabular}{l|ccc|ccc}
		j            &       & 1/2 ($L_2$) &       &       & 3/2 ($L_3$) &       \\
		\hline
		$\mu$        & -1    & 0           & 1     & -1    & 0           & 1     \\
		\hline
		$\uparrow$   & 0.056 & 0.111       & 0.167 & 0.278 & 0.222       & 0.167 \\
		$\downarrow$ & 0.167 & 0.111       & 0.056 & 0.167 & 0.222       & 0.278 \\
		\hline
    unpolarised  & 0.111 & 0.111       &0.111  & 0.222 & 0.222       & 0.222 \\
  \end{tabular}
  \caption{Prefactors $C_{j \mu}^{\uparrow, \downarrow}$. The first two rows are  the weigthing factors for the transitions when the final states are completely (up or down) spin polarised. The third row gives the weighting factors for unpolarised final states. (All per electron).}
\label{tab:xxx}
\end{table}

From here on we focus on the $L_3$ edge and omit the index $j$ for easier readability. Scrutinising Eq.\ref{4} for an incident plane wave,  one notes that $|\varphi_\mu|^2 =|\varphi_{-\mu}|^2$ because the phase factors of both the incident plane wave and the kernel -  $e^{\pm i \alpha}$ -  cancel in the intensity. That means that there will be no difference in the image predicted by Eq.\ref{4} for spin up or for spin down polarisations. The same is true for the intensity in the diffraction plane Eq.\ref{4b}. Essentially, it is not possible to see spin polarisation from single atoms in the TEM without further action. (In the standard EMCD geometry, one uses the interference terms caused by Bragg scattering of the outgoing atomic vortices on the lattice, which is different for spin up and spin down polarisations.)

\begin{figure}
	\centering
		\includegraphics[width=\columnwidth]{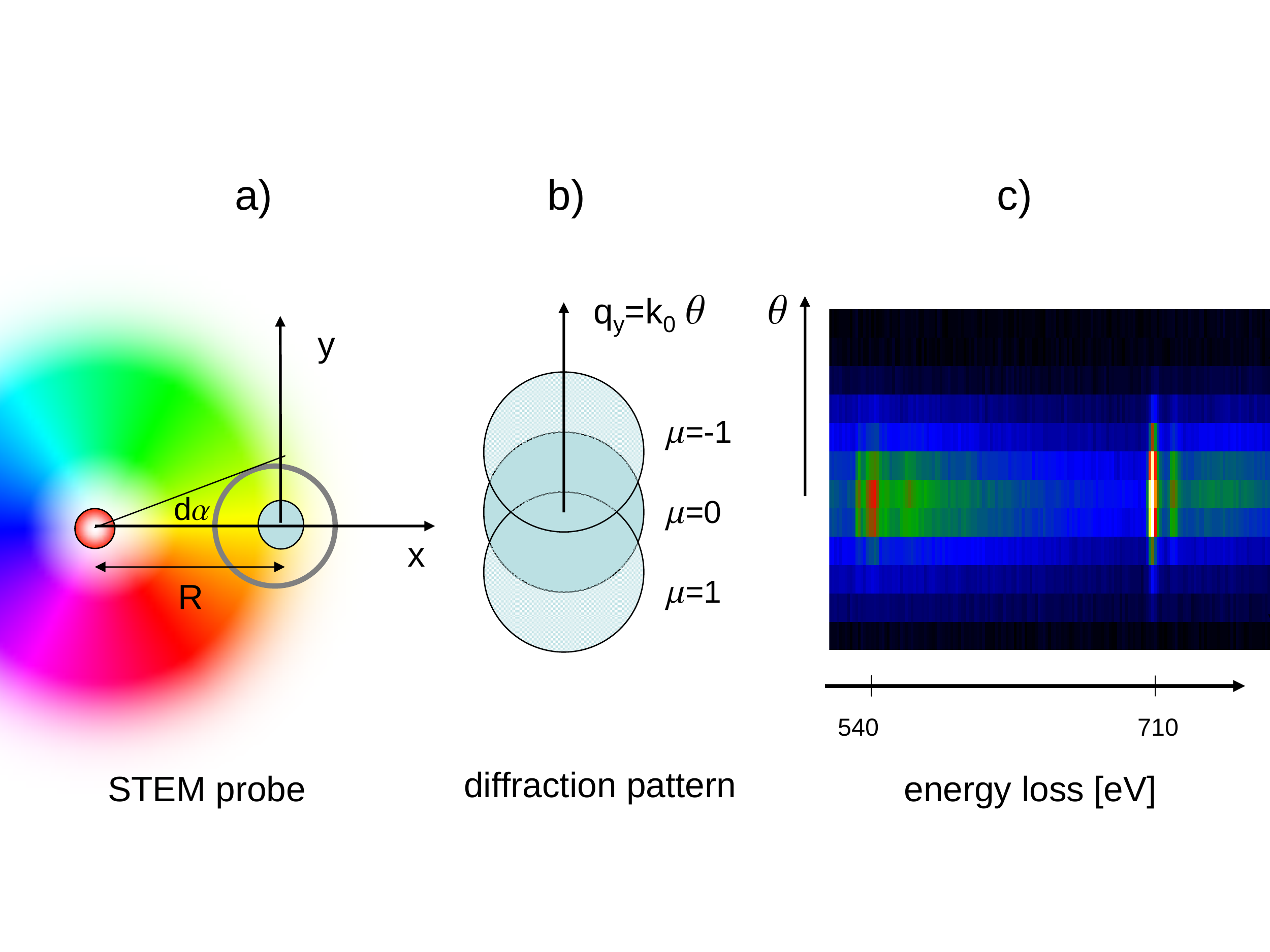}
		\caption{Schematic of the principle of probe-vortex coupling. a) Top view of a  narrow incident Airy disk (gray), focussed at a distance  $R$ in scan direction ($x$) from the atom (red). The scattering kernel $T_E$ is symbolised as a diffuse cloud with a color coded phase (hereafter called rainbow wheel), increasig from blue to red in clockwise rotation for the $\mu=1$ channel. According to Eq.\ref{ramp},  the outgoing wave (gray circle)  has acquired a phase ramp   $d\alpha= \mu y /R$ in the  chiral trasition.  b) The phase ramp translates into a shift of the diffraction disk by $\mu/R$ in direction $q_y$.  c) the $q_x$ extension is squeezed into one pixel on the detector in the $(q,E)$ geometry. Note that the energy loss axis must be  perpendicular to $q_y$, which is proportional to the scattering angle, $q_y=k_0 \theta$. The $(\theta,E)$ map  is a true image taken on magnetite ranging from the oxygen K edge at $\sim 540$~eV to  the Fe L$_{23}$ edge at $\sim 710$~eV. }
	\label{fig:fig1}
\end{figure}

The situation changes  when using a STEM probe instead of a plane incident wave.
We analyse the situation qualitatively before considering numerical simulations. For the explanation of the effect we assume  a   narrow focussed probe given by the Airy function $\mathcal A$ with a diameter  much smaller than the distance $\bf R$ from the atom, essentially so small that that the amplitude of the kernel is almost constant within the probe. (In the numerical simulation this condition is relieved.) Then we can approximate the outgoing wave in channel $\mu$ as 
\begin{equation}
\varphi_\mu({\bf r})=  \psi_\mu({\bf r}-{\bf R}) {\mathcal A}({\bf r}) \approx {\mathcal A}({\bf r})|\psi_\mu({ R})|e^{i \mu \alpha({\bf r})}.
\label{ramp}
\end{equation}
 Note that the outgoing wave is more extended than the Airy disk because of the long range Coulomb coupling force; in Fig.~\ref{fig:fig1}, it is drawn not to scale (even larger for better visibility).  
The azimuth angle within the outgoing disk is $\alpha \doteq \alpha({\bf R})+ \mu y/R$  in the  coordinate system shown in Fig.\ref{fig:fig1}
and we see  that the phase of the outgoing wave depends on the position of the STEM probe and changes sign when going from $\mu=1$ to -1.  Via the shift theorem the diffraction pattern will  be proportional to the Fourier transform of the Airy function shifted in $q_y$ direction by $-1/R,0,1/R$  for the 3  transition channels, that are shifted disk functions $\Pi(\bf q)$
\begin{equation}
\rho_\mu({\bf q},{\bf q}) ={\bar{ C}_{j \mu}}|\psi_\mu({ R})|^2 \Pi({\bf q}+ \mu {\bf \hat{q}}_y /R).
\end{equation}
Having established an observable that has the signature of a particular transition channel the spin polarisation of a single atomic column can be determined.

\section{Numerical simulations}
\begin{figure}
	\centering
		\includegraphics[width= \columnwidth]{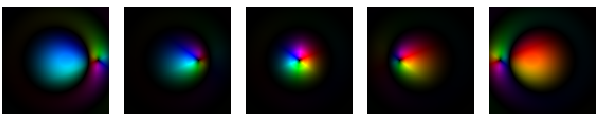}
				\includegraphics[width= \columnwidth]{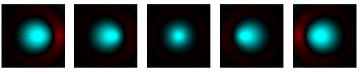}
								\includegraphics[width= \columnwidth]{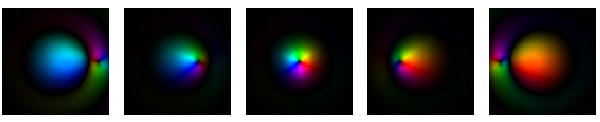}
		\caption{Fe $L_3$ energy filtered real space exit wave functions $\varphi_\mu$ --- Eq.~\ref{phimu} --- of a 100~keV STEM probe of 0.1 nm diameter scanning across a single atom for the 3 dipole allowed transition channels. From top to bottom: transition channels $\mu= -1,0,+1$. From left to right: distance $R$ to the atom -2,-1,0,1,2 at.~units. Brightness codes for intensity of the image, color codes for the phase of the wave function (Color online: rainbow wheel as given in Fig.~\ref{fig:fig1}). When the Airy disk sits on the atom (center column) the outgoing beam is a real vortex with $\mu \in [-1,0,1]$. At a distance phase ramps in the Airy disks develop, changing sign with $R$ and with $\mu$. Each square has a side length of 5 at. units ($\sim$~0.26 nm). }
	\label{fig:Scanseries}
\end{figure}

Simulations were performed for an incident probe of 100~keV. 
First we construct the outgoing signal for each of the 3 transition channels, $\rho_\mu({\bf r,r})$, shown in Fig.\ref{fig:Scanseries}.
These are $L_3$ energy filtered images of a STEM probe of 0.1 nm diameter scanning across a single atom. From top to bottom are the  transition channels $\mu= -1,0,+1$. From left to right the distance $R$ to the atom is -2,-1,0,1,2 at.~units. Note that the STEM probe is always in the center of the images. The side structures at distances of $R=\pm 2$~at.u. are signals from the second maximum of the Airy disk that coincides here with the atom centre.  Brightness codes intensity of the image, color codes for the phase of the wave function. (Color online: rainbow colors from $-\pi$ to $\pi$). When the Airy disk sits on the atom the outgoing beam is a true atomic vortex with topological charge $\mu \in [-1,0,1]$. At larger distance phase ramps in the Airy disks develop, visible as continuous color variations. They  change sign when crossing the atom centre and are opposite for  $\mu=\pm 1$. Each square has a side length of 5 at.~units (0.26 nm).

\begin{figure}
	\centering
		\includegraphics[width= \columnwidth]{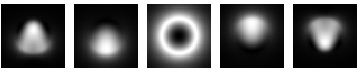}
				\includegraphics[width= \columnwidth]{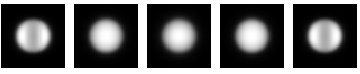}
								\includegraphics[width= \columnwidth]{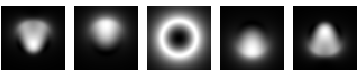}
		\caption{Fe $L_3$ energy filtered diffraction patterns $\rho({\bf q},{\bf q})_\mu$ --- Eq.~\ref{4bb} --- of a 100~keV STEM probe of 0.1 nm diameter scanning across a single atom, corresponding to Fig.\ref{fig:Scanseries}. From top to bottom: transition channels $\mu= -1,0,+1$. From left to right: distance $R$ to the atom -2,-1,0,1,2 at. units.  Note the shift of the patterns for $\mu=\pm 1$ in vertical direction $q_y$ as predicted in Fig.\ref{fig:fig1}b, depending on the position of the probe, and the inversion of shifts with change of sign. The $\mu=0$ channel does not show any shift because there is no phase ramp in the image.  Convergence angle 18~mrad. The images have a side length of  $\pm 50$~mrad. }
	\label{fig:DPseries}
\end{figure}

These phase ramps are responsible for the corresponding shifts of the diffraction patterns shown in Fig.\ref{fig:DPseries}. The shift of the patterns  in vertical direction ($q_y$) is opposite for $\mu=\pm 1$ and depends on the position of the probe. The $\mu=0$ channel does not show any shift because it lacks a phase ramp in the image. The patterns are smeared by convolution of the incident probe disk with the inelastic scattering kernel that has an extension of $\delta \theta \sim \Delta E/(2 E_0) \approx 3.5$~mrad.

Since the $\mu = \pm 1$ channels contribute differently for spin up/spin down polarisation via the coefficients $C_{j\mu}$, Eq.\ref{Cj}, the diffraction patterns will be different for these two cases. (Note that this is not the case in the filtered real space image: there, the difference is only in the phase, not in the intensity distribution).

Monitoring energy filtered diffraction patterns of a scanned probe means measuring a multidimensional data cube because background subtraction and multiple scattering deconvolution of energy loss spectra require a range of losses. Such data has two dimensions (qx,qy) in the reciprocal space, one in the energy-loss (eV), and one (x) or two (x,y) in the real space, depending on the scanning pattern. This creates huge data files  - a scan over 1 elementary cell in magnetite with 0.02~nm step width with 256$^2$ pixels in the diffraction pattern would give $\sim 60$~Mb per energy channel, approaching $\sim 10$~GB for a whole spectrum -  and is impractical. One can however exploit a remarkable feature apparent in  the simulations: The diffraction patterns show only  asymmetry with respect to the coordinate $q_y$, which is the Fourier transformed variable of $y$. An experimental setup could then discard the $q_x$ variable by integration without  information loss. This is exactly what is realised in the $(q,E)$ geometry. There, the $q_x$ axis is squeezed onto one pixel of the specrometre by compressive lenses whereas the $q_y$ axis is retained, being projected on the detector perpendicular to the  energy loss axis. This is sketched in Fig.\ref{fig:fig1}c which is a $(q_y,E)$ data set for a fixed positon $R$ . Selecting the $L_3$ white line from the whole $(q_y,E,R)$ data, a reduced subset with axes $(R,q_y)$ is obtained.  The scan direction must be perpendicular to $y$ in the coordinate system of Fig.\ref{fig:fig1}. Contrary to XMCD or EMCD, the new technique operates on a single white line only (the stronger $L_3$ line here). This is important because often the $L_2$ edge is too faint to obtain sensible results. (It should be noted that for  the separation of spin and orbital moments both $L_2$ and $L_3$ edges are needed~\cite{RuszPRB07,CalmelsPRB07}.

\begin{figure}
\begin{minipage}[t]{\linewidth}
	\centering
\includegraphics[width= \columnwidth]{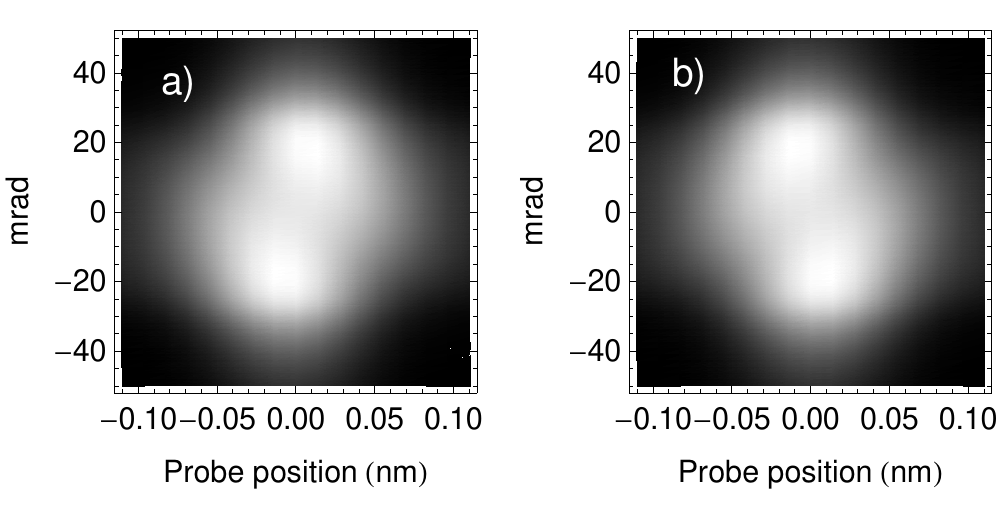}	
	\end{minipage}
	\hspace{0.2 cm}
	\begin{minipage}[t]{\linewidth}
	\centering
\includegraphics[width= \columnwidth]{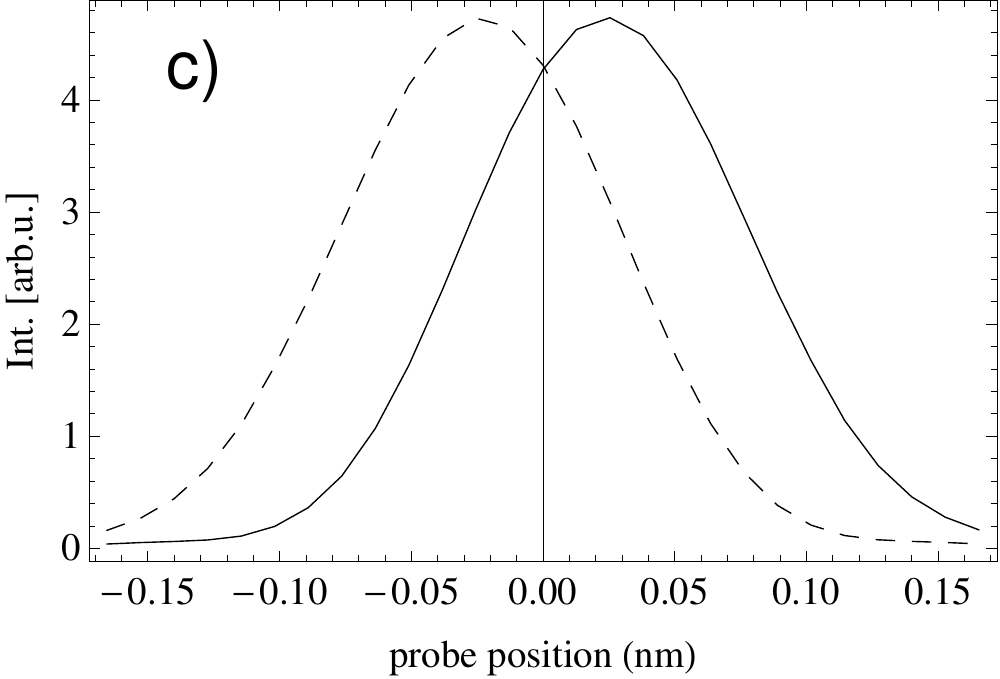}
\end{minipage}
\caption{Fe $L_3$ energy filtered $(R,q_y)$ signal distribution of a STEM probe of 0.1 nm diameter scanning across a single atom, corresponing to Fig.\ref{fig:Scanseries}. a: spin up polarisation of the atom; b: spin down polarisation. The top/bottom asymmetry is well visible and can be used to determine the spin polarisation. 
c: Full line: Intensity integrated over the upper detector half (20 to 50 mrad) line scan  of a STEM probe of 0.1 nm diameter  across a single atom, corresponding to a). Dashed line: same for the lower detector half (-50 to -20 mrad).  }
\label{fig:Scantotal}
\end{figure}

These subsets can be constructed from the previous results, simply by integrating the intensity of Fig.\ref{fig:DPseries} over the "`squeezed"' variable $q_x$ and summing over the 3 transition channels in Eq.\ref{4b}. The result is shown in Fig.\ref{fig:Scantotal} for spin up and spin down configurations. The asymmetry of the intensity distribution with respect to the atom position at $R=0$ is indicative for spin polarised transitions. 

To enhance  the asymmetry Fig.\ref{fig:Scantotal} suggests to avoid the central part around $q_y=0$ which adds only spin-insensitive intensity thus increasing the noise level. Integration of the density matrix Eq.\ref{4bb} over the scattering angle in the top and bottom parts   yields two scans over a single atom
\begin{eqnarray}
\rho_+(R)&=&\int dq_x \int_{q_{y1}}^{q_{y2}}\rho({\bf q,q})\, dq_y \cr
\rho_-(R)&=&\int dq_x \int_{-q_{y2}}^{-q_{y1}}\rho({\bf q,q})\, dq_y
\end{eqnarray}
where $q_y=k_0 \theta$.
This is shown in Fig.\ref{fig:Scantotal}c for $\theta_{y1}=20$~mrad, $\theta_{y2}=50$~mrad. The difference in position between the  maxima in the two scans is $\sim$~0.06~nm, indicating that  Cs corrected machines and extreme stability are needed to see the effect. Even so, the signal will be very faint, such that noise will tend to override the effect. Dynamical diffraction of the incident and the outgoing electron on the lattice and remaining aberrations of the probe forming lens and the spectrometer will also complicate the situation. Quantitative spin detection will therefore need elaborate calculations.

\section{Experimental evidence}
We have performed a feasibility experiment at the Daresbury SuperSTEM facility using a NION UltraSTEM100 microscope\cite{NION}. The instrument is a dedicated, aberration corrected STEM operated at 100 kV with a cold-FEG emitter. Its 3rd generation C3/C5 aberration corrector allows a typical probe size of 0.1~nm at a beam current of 30~pA. EELS spectrum imaging at atomic resolution is done with a DigiScan2 scanning unit in combination with a Gatan Enfina spectrometer. 
The difficulty of the experiment stems from the extreme demand on stability, the rather long dwell times causing beam damage, remaining aberrations, and the low count rates. Spectra were collected during 0.2~s per position, corresponding to a dose of $\sim$ 5. 10$^7$ electrons focussed in the STEM spot. 

\begin{figure}
	\centering
\includegraphics[width= 0.75\columnwidth]{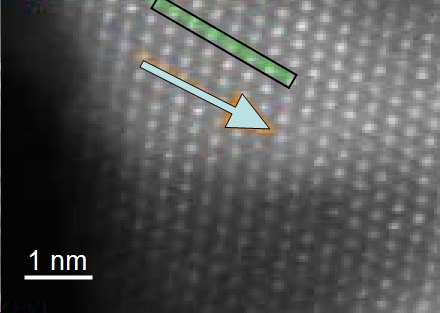}
		\caption{HAADF image of the border of the magnetite nanoparticle in [111] zone axis orientation. The scan line is indicated as a box, the axis of energy dispersion (arrow) is almost perfectly parallel to the scan direction, thus $q_y || y$. The bright dots are the A columns containing 3 Fe atoms per unit cell in this projection. Their projected distance is 0.33~nm.}
	\label{fig:HAADF}
\end{figure}

\begin{figure}
	\begin{minipage}[b]{\linewidth}
		\centering
\includegraphics[width= \columnwidth]{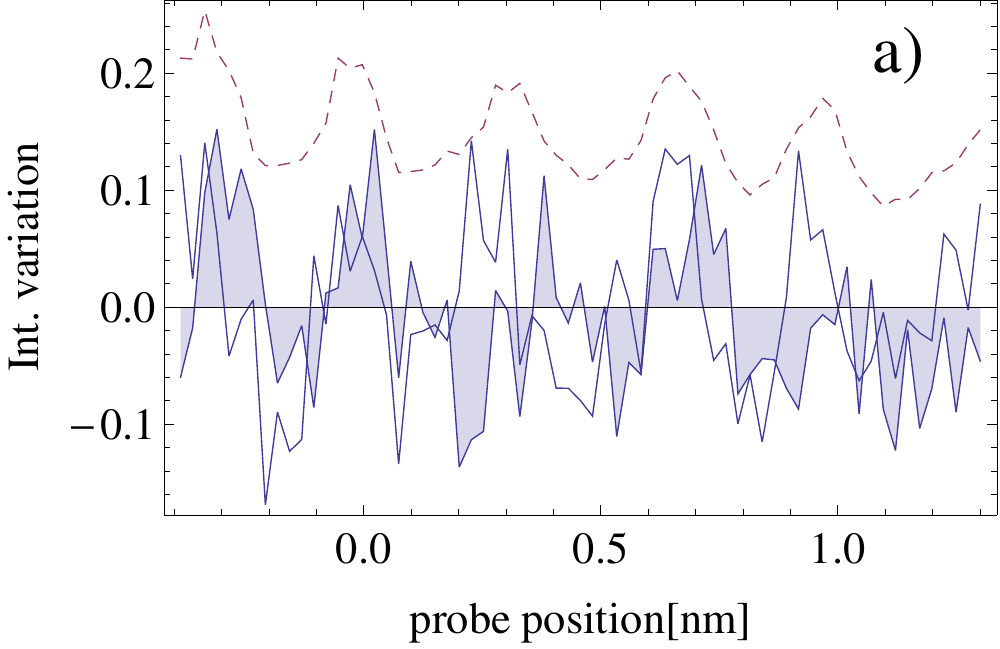}	
\end{minipage}
\vspace{0.2 cm}
\begin{minipage}[b]{\linewidth}
	\centering
\includegraphics[width= \columnwidth]{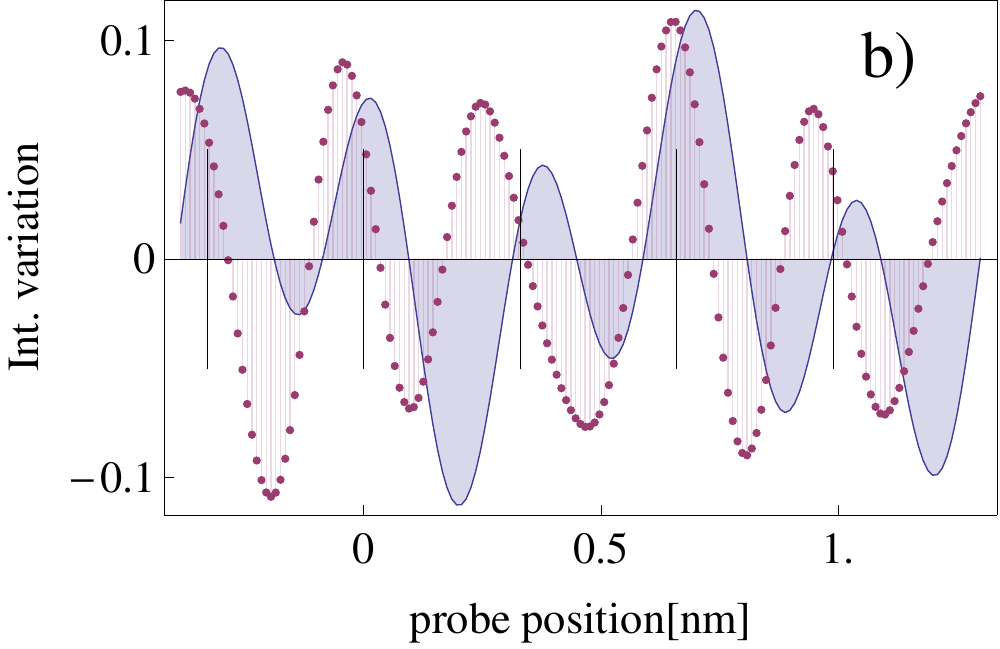}
	\end{minipage}
	\caption{a: Scan of the Fe $L_3$ white line signal after standard background subtraction, drift correction and removal of the continuum. The filled  curve shows the scan using the upper detector half (20 to 50 mrad), the empty curve is a scan using the lower detector half (-50 to -20 mrad). The HAADF signal is superposed (dashed) with maxima indicating  the positions of the A columns. The B columns are visible as  faint subsidiary peaks. b: 	Same scans after  Fourier filtering. Atom positions (MAADF maxima) are marked with vertical lines.}
		\label{fig:BSprocessed}
\end{figure}

We investigated  a  platelet-like magnetite nanoparticle of $\sim$~15 nm diameter, Fig.\ref{fig:HAADF}. 
The thickness in this region was between 5 and 10~nm, resembling  as  close as possible  the single atomic row model. The STEM probe was scanned over a line of atoms marked  in fig.\ref{fig:HAADF} as a rectangle. Magnetite is an inverted cubic spinel $XY_2O_4$ with $Fe^{3+}$ ions at tetrahedral $X$ sites. The octahedral $Y$ sites are randomly occupied by $Fe^{2+}$ and  $Fe^{3+}$ ions. $X$ and $Y$ sites are antiferromagnetically coupled. In (111) zone axis there are two types of Fe columns (hereafter called A and B). A contains two $X$ sites and one $Y$ site per elementary cell, B contains only one  $Y$ site. The B columns are not visible in the HAADF image Fig.\ref{fig:HAADF} because of lower scattering strength and dynamical diffraction, confirmed by multislice simulations. However, very faint side maxima from the B columns can be seen in 
Fig.~\ref{fig:BSprocessed}a. Shown there are the   top and bottom profiles of the scan integrated from 20 to 50~mrad after standard background subtraction, drift correction  and usual removal of the continuum signal beneath the white line. The post-edge continuum, often used in standard EMCD for normalising, did not show periodic variations. Therefore it was not necessary to correct for, the more so as this would have induced additional  noise. A  shift of the traces with respect to the atomic positions can already be guessed although the noise is almost overriding the signal.
The number of electrons collected in the $L_3$ edge was $\sim$~250 in each half detector, causing a theoretical  shot noise level of $3 \sigma=47$. Pre-edge extrapolation, instability and other error sources add noise such that the $3 \sigma$ relative error amounts to $\sim 20 - 30\%$. 

In order to demonstrate  the predicted effect qualitatively a Fourier analysis was performed on the two scans, retaining only coefficients up to lattice periodicity. 
The result is shown in Fig.~\ref{fig:BSprocessed}b. As predicted in Fig.\ref{fig:Scantotal} the maxima are  shifted to both sides of the atom centres that are marked by vertical lines. The average distance between the top and bottom scan maxima is 0.076~nm. One observes a rather large shift at the 0.33~nm position that is  caused by some irregularity (probably sudden drift). Excluding this value, the average difference between the respective maxima in the two scans is 0.65~nm, in good agreement with the simulational result of 0.6~nm seen in Fig.\ref{fig:Scantotal}b.
Despite  the crude approximations and the simple model  the agreement is surprisingly  good. More accurate  models can be devised, but this is beyond the scope of the present paper as  noise, drift and beam damage  pose narrow  limits on the interpretation of the data.  
More elaborate experiments  with  ultrathin magnetic specimens must be performed in order to confirm the present findings.
 
 \section{Potential applications}
 \subsection{Detection of magnetic order}
\begin{figure}[htb]
	\centering
	\begin{minipage}[b]{\linewidth}
		\centering
\includegraphics[width=\columnwidth]{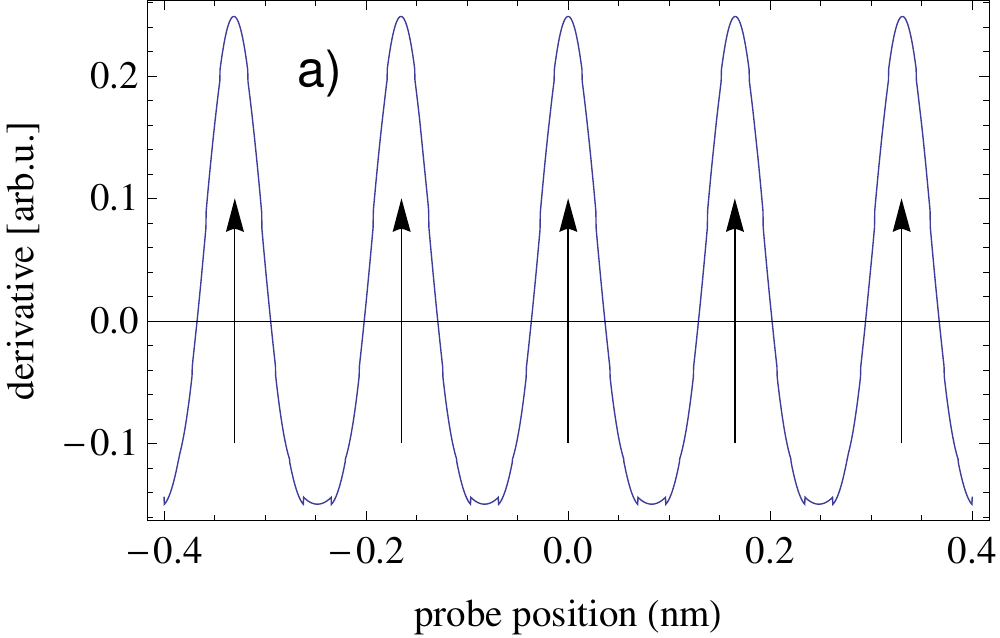}	
\end{minipage}
\vspace{0.4 cm}
	\begin{minipage}[b]{\linewidth}
		\centering
\includegraphics[width= \columnwidth]{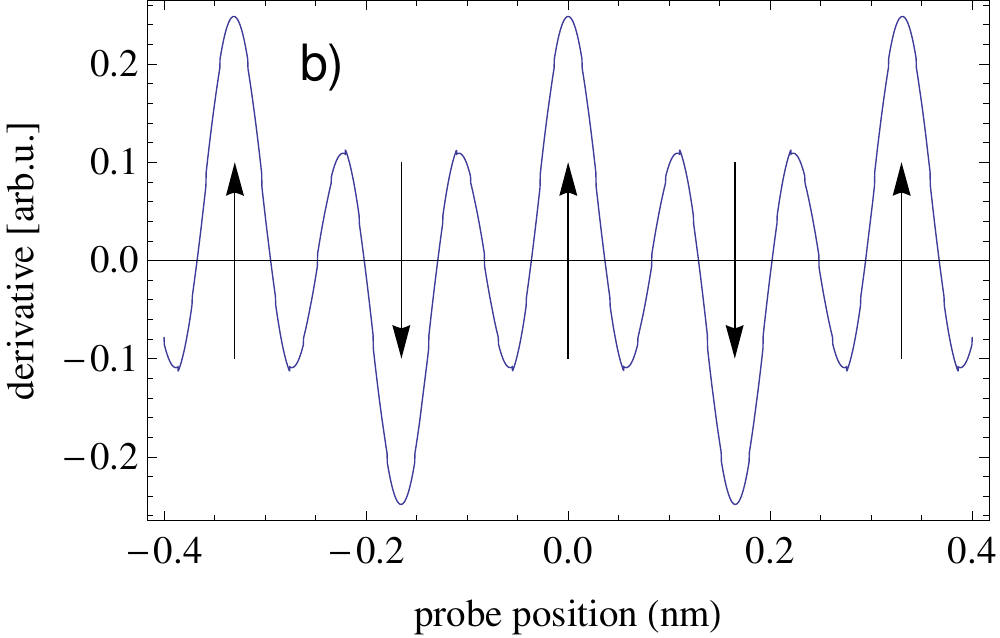}	
\end{minipage}
\vspace{0.2 cm}	
\begin{minipage}[b]{\linewidth}
		\centering
\includegraphics[width= \columnwidth]{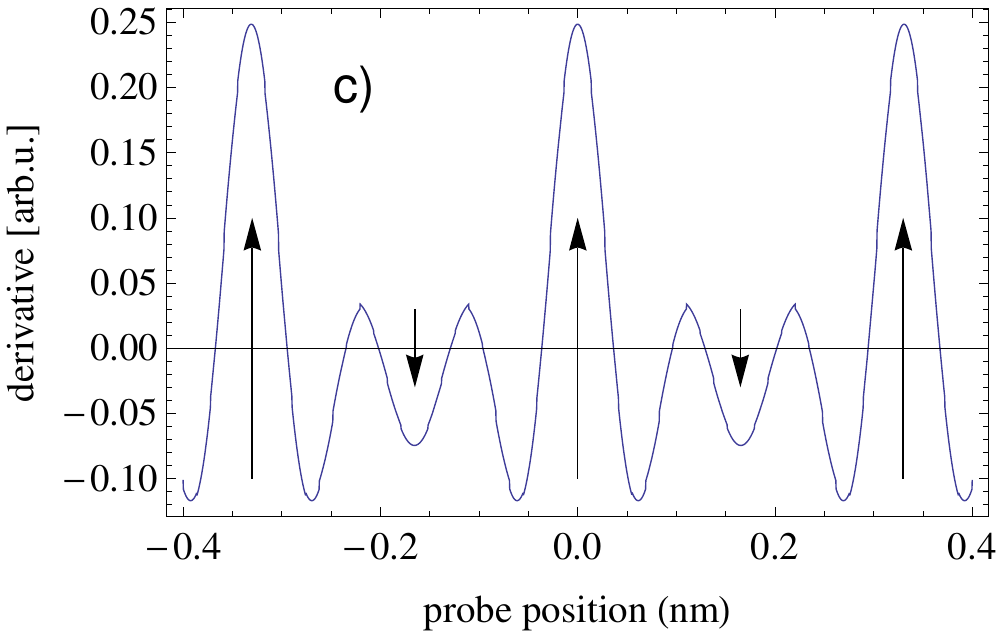}	
\end{minipage}
\vspace{0.2 cm}
		\caption{Hypothetical line scans showing the first derivative $d(\rho_+(R) - \rho_-(R)/dr$ over an atomic row as in Fig.~\ref{fig:HAADF} for a: ferromagnetic, b: antiferromagnetic and c: ferrimagnetic ordering. c) was calculated with a relative strength of -30 \% for the spin down moments, thus taking into account the strong channeling at the A columns.}
	\label{fig:MagneticOrder}
\end{figure}

As described above and shown in Fig.\ref{fig:Scantotal}, the top/bottom asymmetry in the $(R,q_y)$ data set is caused by the spin polarisation on the atomic site. That provides a method for spin mapping.
A straightforward way to do so is to take the first derivative of the difference (top-bottom) line scans 
$(\rho'_+-\rho'_-)$. Since the slope of the difference signal is strongest at the atomic sites, this scan gives directly the sign and position of the magnetic moment.
Here  we assume a hypothetical system with the same lattice constant as magnetite, also in (111) zone axis orientation, to make connection to real systems.
Fig.\ref{fig:MagneticOrder} shows the derivative $(\rho_+-\rho_-)'(R)$ along a line scan as in Fig.\ref{fig:HAADF}, for different assumptions of the magnetic ordering (ferromagnetic, antiferromagnetic, and ferrimagnetic).  To take account of the dechanneling and defocussing throughout the specimen the data was convolved with a 0.1 nm broadening function. The last panel resembles magnetite, in fact.
 In (111) zone axis projection the A columns contain 2 atoms spin up and on atom spin down, and the B columns contain one atom spin down, oriented along the $z$ axis in the magnetic field of the objective lens. On the other hand, the channeling is stronger on the deeper  potential (A columns containing 3 atoms as compared to 1 on B) so the beam will see more from the A columns than from the B columns. This will  weigh the A columns stronger than the B ones. The exact weighting factor is impossible to obtain without solving the dynamical equation for the propagator in the lattice, but a weighting coefficient  between -50\% and -25\% for the B columns is  reasonable. Here, -30\% were assumed for the simulation. Theoretically, it should be possible to perform such scans not only along lines but over areas. This should give atom-resolved maps of the element specific magnetic moments and the magnetic ordering. But before establishing such an analytical technique the   problems related to  noise, stability, aberrations and dynamical diffraction must be solved.
\subsection{Free electrons with angular momentum}
\begin{figure}
	\centering
\includegraphics[width= \columnwidth]{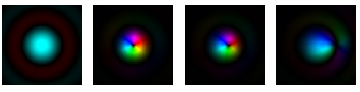}
		\caption{The phase and amplitude of the incident Airy disk of 0.1~nm diameter (left) and for  the scattered wave from the $\mu=1$ transition. The atom is displaced from the beam center to the left by 0, 0.01, and 0.05~nm. The images have a side length of 0.2~nm.}
	\label{fig:VortexPhase}
\end{figure}

\begin{figure}
	\centering
\includegraphics[width= \columnwidth]{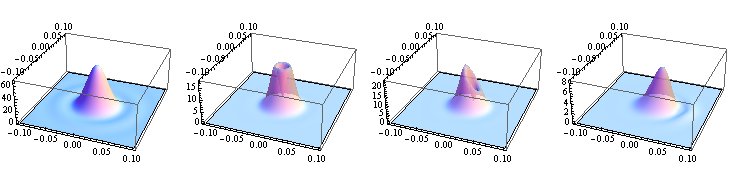}
		\caption{Intensity of the incident Airy disk (left) and $Tr[\rho]$ for  the scattered wave (mixed state from 3 transition channels) for atom displacements as in Fig.~\ref{fig:VortexPhase}.}
	\label{fig:Vortex3D}
\end{figure}

\begin{figure}
	\centering
		\begin{minipage}[b]{\linewidth}
		\centering
\includegraphics[width= \columnwidth]{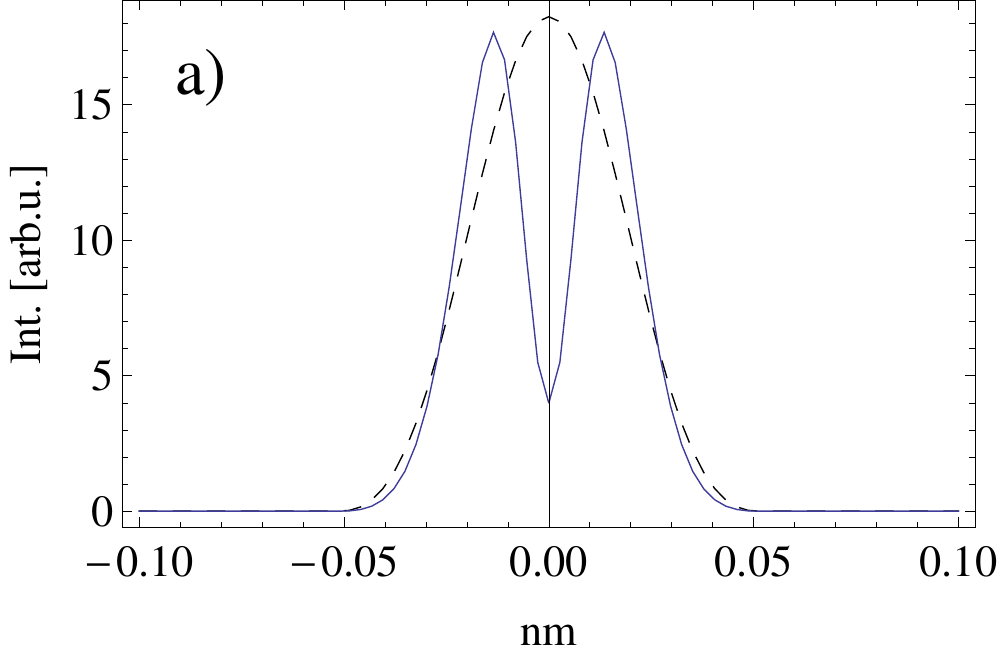}	
\end{minipage}
\hspace{0.2 cm}
	\begin{minipage}[b]{\linewidth}
		\centering
\includegraphics[width= \columnwidth]{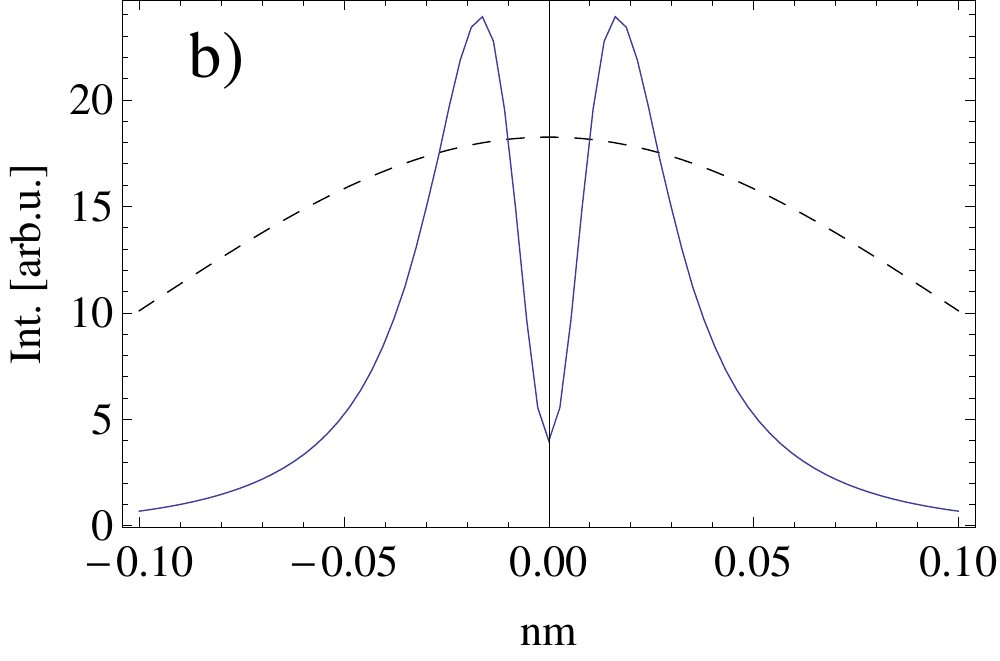}	
\end{minipage}
		\caption{a) Radial profiles of the incident one-Angstrom probe (dashed)  and the outgoing electron density of the mixed state when the atom is exactly centred on the beam. b) The same for an incident probe of 5 Angstrom diameter. }
	\label{fig:VortexProfile}
\end{figure}

If an Angstrom sized spot is focussed  exactly at the centre of an atom in a ferromagnet the scattered electron has acquired orbital momentum. It is important to note that it is in a mixed state with contributions from the three transition channels $\mu \in [-1,1]$, each creating a pure vortex state with topological charge $\mu$. 
The expectation value of the angular momentum can be calculated from the coefficients $C_{j\mu}$ given in Eq.~\ref{Cj}:
$$
\expect{{\mathcal L}_z}=\frac{Tr_\mu[{\mathcal L}_z \rho]}{Tr_\mu[\rho]} \in [ -0.167 \hbar, 0.167 \hbar] 
$$
depending on the spin polarisation of the atom. This is a unique and simple method to create free electrons with orbital momentum, although the efficiency and the available momentum is probably too low to manipulate nanoparticles or even atoms by the torque. Another problem is the precise positioning of the probe on the atom.  Fig.~\ref{fig:VortexPhase} shows the phase and amplitude of the incident beam of one Angstrom diameter and the scattered beam for the $\mu=1$ transition when the atom is in the center, 0.01 nm, and 0.05 nm sideways. The vortex structure disappears rapidly, also visible in the intensity distribution Fig.~\ref{fig:Vortex3D}. Here, the total electron density $Tr_\mu[\rho]$ is shown. The central dip, characteristic for the topological charge, remains at a deviation of 0.01~nm but has disappeared for 0.05~nm. Fig.~\ref{fig:VortexProfile}a compares the incident Airy disk with the radial profile of the outgoing mixed state. The scattered state is even narrower than the Airy disk. The high sensitivity of the outgoing vortex state to the probe position could be used for a more direct method of spin mapping with sub-atomic resolution than described above, e.g. with a vortex filter such as a holographic mask. 

We simulated also the scattered state when the incident beam is broader (0.5~nm diameter). The result is shown in Fig,~\ref{fig:VortexProfile}b. The outgoing beam is broader than in case a) but still much narrower than the incident Airy disk, and is again a superposition of electron vortices. The not so surprising consequence is that  a thin ferromagnetic foil in an electron beam creates free electrons carrying angular momentum after energy filtering.
This shows that vortices always have been there in EELS experiments on magnetic materials.

\section{Conclusion}
The coupling of an Angstrom-sized electron  probe to a spin polarised transition creates a mixed state that contains electron vortices with non-zero orbital momentum.  These states   break the symmetry of  the scattering distribution in the far field in a way  characteristic for the chirality of the transition, a fact that can be used for the imaging of electron spins in real space with sub-Angstrom resolution. 
A tentative experiment on a magnetite nanoparticle shows  the expected asymmetry.   

Apart from  probing the local magnetic ordering, important for a number of technologically promising materials such as Heusler alloys, the proposed method bears promise for the mapping of spin polarisations of single atomic columns, be that in the vicinity of interfaces, magnetically dead layers, or magnetic core-shell structures.

The creation of free electrons carrying angular momentum  is theoretically  feasible via spin polarised electronic transitions. This works  even for relatively broad incident beams passing a thin ferromagnetic foil. 

{\bf Acknowledgements:} The authors thank  Andrew Bleloch, Stefan L{\"o}ffler and Peter Nellist for fruitful discussions and suggestions.
P.S. acknowledges financial support from the Austrian Science Fund, Project I543-N20.
The support of the EPSRC for the SuperSTEM facility is gratefully acknowledged.
J.V. acknowledges support from the European Research Council under the 7th Framework Program (FP7), ERC grant Nr.~246791 - COUNTATOMS and ERC Starting Grant 278510 - VORTEX. 


\end{document}